\documentclass[reprint,prb,aps,longbibliography,floatfix]{revtex4-2}
\usepackage{graphicx}
\usepackage{amsmath,amssymb}
\usepackage{xcolor}

\usepackage[
  colorlinks=true,
  linkcolor=blue,
  citecolor=magenta,
  urlcolor=magenta,
  hypertexnames=false
]{hyperref}

\newcommand{\tcell}[2]{\parbox[t]{#1}{\raggedright #2}}

\begin{document}
\title{Magnetic-field-induced detwinning of orientational domains in altermagnet $\alpha$-MnTe: A single-crystal neutron diffraction study}

\author{Feihao Pan$^{1}$}
\author{Yishui Zhou$^{1}$}
\author{Sabreen Hammouda$^{1}$}
\author{Yinghao Zhu$^{1}$}
\author{Soohyeon Shin$^{1}$}

\author{Karin Schmalzl$^{2}$}
\author{Wolfgang Schmidt$^{2}$}

\author{Eric Ressouche$^{3}$}
\author{Oscar Fabelo$^{4}$}

\author{Yixi Su$^{1}$}
\email[Contact author: ]{y.su@fz-juelich.de}

\affiliation{$^{1}$J\"ulich Centre for Neutron Science (JCNS) at the Heinz Maier-Leibnitz Zentrum (MLZ), Forschungszentrum J\"ulich, Lichtenbergstrasse 1, D-85747 Garching, Germany}
\affiliation{$^{2}$J\"ulich Centre for Neutron Science (JCNS) at ILL, Forschungszentrum J\"ulich, F-38000 Grenoble, France}
\affiliation{$^{3}$IRIG, CEA and Université Grenoble Alpes, CEA Grenoble, F-38054 Grenoble, France}
\affiliation{$^{4}$Institut Laue-Langevin, 71 avenue des Martyrs, 38042 Grenoble Cedex 9, France}

\date{\today}

\begin{abstract}
Altermagnets combine compensated magnetic order with momentum-dependent spin splitting, which makes them promising for future spintronic and magnonic applications, but their macroscopic responses can be strongly reduced by domain averaging.
In hexagonal $\alpha$-MnTe, three symmetry-related in-plane N\'eel-vector orientations and their time-reversal counterparts form six magnetic domain states, complicating both magnetic structure determination and measurements of the chiral magnon response.
Using single-crystal neutron diffraction, we show that an in-plane field applied along the $[-1,1,0]$ direction produces a common magnetic domain orientation that saturates near $\mu_0H=2$~T.
A constrained refinement of 61 nuclear and magnetic reflections collected at 2~K and 2~T yields $\mathbf{M}\parallel[1,1,0]$, an ordered moment of $4.62(8)~\mu_\mathrm{B}$ per Mn, and an orientational domain population at $r_A:r_B:r_C=0.97:0.03(5):0$ that is compatible with a nearly complete alignment of orientational domains.
Furthermore, an abnormal behaviour in the temperature dependence of the selected reflections is observed in 60--88~K, its microscopic origin is discussed within the context of possible intrinsic piezomagnetic or magnetoelastic effects.
Because unpolarized diffraction neither distinguishes domain repopulation from continuous N\'eel-vector rotation nor resolves time-reversal antiphase domains, the field-prepared state is a near-single-orientation state rather than a truly single-domain state.
This non-contact domain control provides a practical complement to strain and field-cooling approaches for preparing well-defined magnetic states in bulk $\alpha$-MnTe.
\end{abstract}

\maketitle

\begin{figure*}[t!]
    \includegraphics[width=0.92\linewidth]{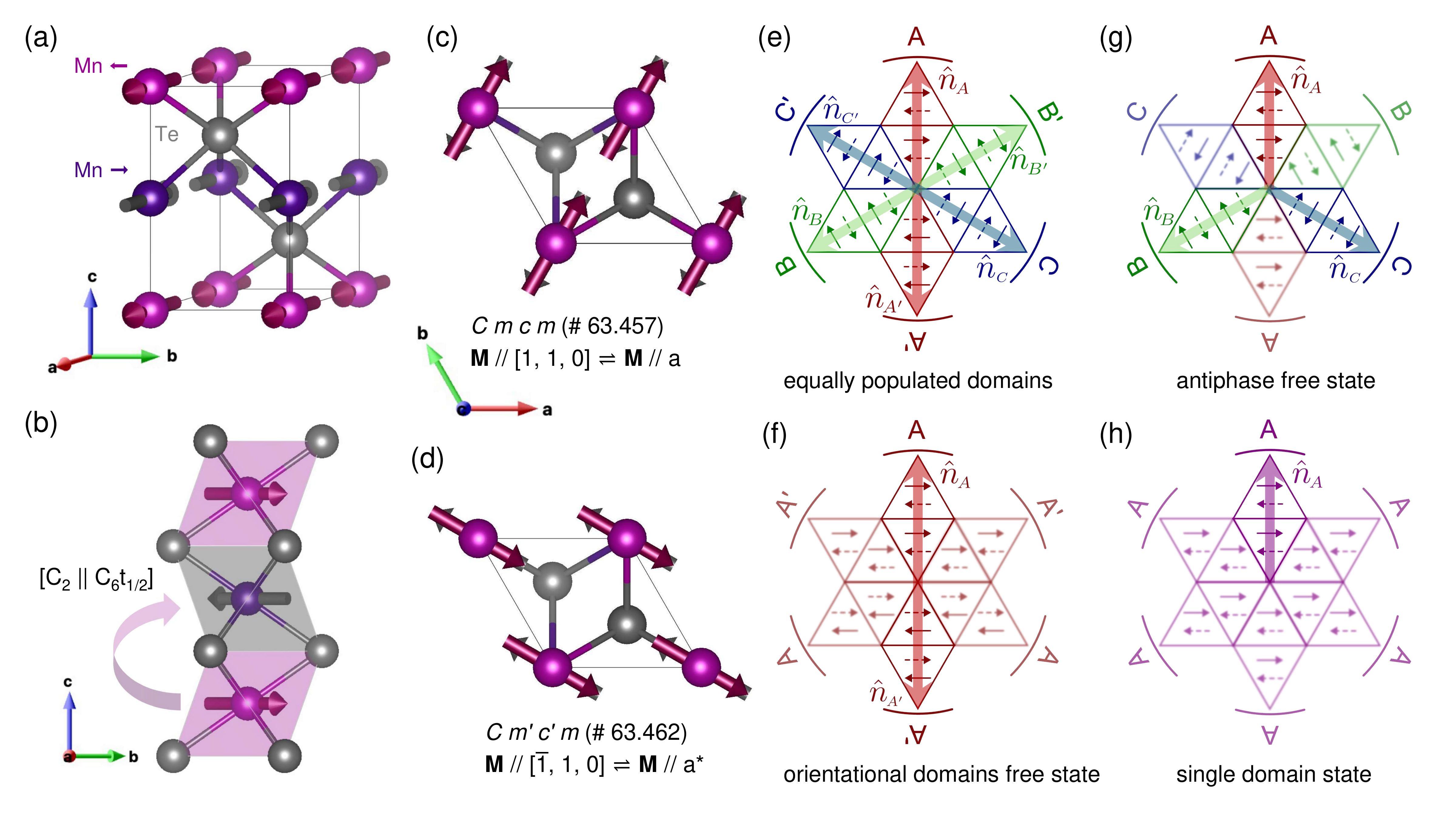}
    \caption{
    \textbf{(a)} Crystal structure of $\alpha$-MnTe, with the two opposite-spin Mn sublayers and the Te ligands.
    \textbf{(b)} Illustration of the $[C_2 \parallel C_6 t_{1/2}]$ spin-space-group operation connecting the opposite-spin Mn layers.
    \textbf{(c), (d)} Magnetic structures for the two in-plane moment orientations: $[1,1,0]$ ($Cmcm$, No.~63.457) and $[\bar{1},1,0]$ ($Cm'c'm$, No.~63.462).
    \textbf{(e)} Six-domain state with equal orientational and antiphase populations. The solid arrow refer to the moment direction in one layer, while the dashed arrow refer to the moment direction in the neighboring layer.
    \textbf{(f)} Single-orientation state in which antiphase partners may coexist.
    \textbf{(g)} Antiphase-selected state in which all three orientational variants may coexist.
    \textbf{(h)} Fully single-domain state.
    }
    \label{fig:fig1}
\end{figure*}

\begin{table*}[t!]
\caption{Comparison of magnetic-domain and magnetic-state control in $\alpha$-MnTe. ND, neutron diffraction; INS, inelastic neutron scattering; PND, polarized neutron diffraction; AHE, anomalous Hall effect. ARPES, angle-resolved photoemission spectroscopy; NMR, nuclear magnetic resonance.}
\label{tab:comparison}
\begin{ruledtabular}
\scriptsize
\setlength{\tabcolsep}{3pt}
\begin{tabular}{lllll}
System & Experimental method & Detwin/control method & Domain or order parameter & Moment direction/Effect \\
\hline
Bulk (this work) & \tcell{2.8cm}{Unpolarized single-crystal ND} & \tcell{3.1cm}{In-plane $\mathbf{H}\parallel[\bar{1},1,0]$} & \tcell{3.0cm}{Orientational} & \tcell{3.8cm}{$[1,1,0]$ ($\sim 97\%$ imbalance)} \\
Thin film~\cite{kriegner_magnetic_2017} & \tcell{2.8cm}{ND} & \tcell{3.1cm}{In-plane $\mathbf{H}$} & \tcell{3.0cm}{Orientational} & \tcell{3.8cm}{$[1,\bar{1},0]$ family; \\ strain tunes the threshold} \\
Bulk~\cite{liu_strain-tunable_2026} & \tcell{2.8cm}{ND and transport (AHE)} & \tcell{3.1cm}{Uniaxial compression \\ $\parallel[1,1,0]$} & \tcell{3.0cm}{Orientational} & \tcell{3.8cm}{$[1,\bar{1},0]$ family} \\
Bulk~\cite{liu_switchable_2026,liu_altermagnetic_order_2026} & \tcell{2.8cm}{PND, polarized INS} & \tcell{3.1cm}{$c$-axis and oblique-field cooling} & \tcell{3.0cm}{Antiphase plus partial orientational selection} & \tcell{3.8cm}{Switchable order and magnon chirality} \\
Li-doped bulk~\cite{moseley_doping_2022,yumnam_magnon_2024,zhang_li-doped_2025} & \tcell{2.8cm}{ND, INS, ARPES} & \tcell{3.1cm}{Chemical substitution} & \tcell{3.0cm}{Orientational/ \\ Easy-axis reorientation} & \tcell{3.8cm}{From in-plane to $[0,0,1]$} \\
Bulk~\cite{duan_magnetostriction_2026,weber_magnetic_2026,huynh_intrinsic_2026} & \tcell{2.8cm}{Magnetostriction, INS, NMR} & \tcell{3.1cm}{In-plane magnetic field} & \tcell{3.0cm}{Depopulation plus continuous rotation} & \tcell{3.8cm}{Field-perpendicular alignment} \\
\end{tabular}
\end{ruledtabular}
\end{table*}

\section{Introduction}

Altermagnet is a recently established class of compensated collinear magnets in which crystal rotations connect opposite-spin sublattices, allowing momentum-dependent spin splitting without a net magnetization~\cite{smejkal_beyond_2022,mcclarty_landau_2024,chen_unconventional_2025,jungwirth_2025,jungwirth_altermagnetism_2025}.
Its broken time-reversal symmetry permits responses such as the anomalous Hall effect and non-relativistic chiral-magnon splitting in selected magnetic configurations~\cite{smejkal_crystal_2020,smejkal_emerging_2022,bai_altermagnetism:_2024,dai_research_2026}.
These properties make altermagnets promising for spintronics and magnonics devices by combining spin-polarized functionality with negligible stray fields.

Among various candidate materials, hexagonal $\alpha$-MnTe (space group $P6_3/mmc$) has become a widely studied prototype of $g$-wave altermagnets~\cite{gonzalez_betancourt_spontaneous_2023,lee_broken_2024,krempasky_altermagnetic_2024,kluczyk_coexistence_2024,amin_nanoscale_2024}.
Neutron diffraction established an A-type antiferromagnetic structure with propagation vector $\mathbf{k}_m=(0,0,0)$, comprising ferromagnetic Mn layers stacked antiferromagnetically along the $c$ axis~\cite{kunitomi_neutron_1964,efrem_dsa_low-temperature_2005} (Fig.~\ref{fig:fig1}\textbf{(a)}).
Subsequent non-polarised inelastic neutron scattering investigations determined the dominant exchange interactions~\cite{szuszkiewicz_spin-wave_2006} and more recently revealed split magnon bands that are consistent with the altermagnet theory~\cite{liu_chiral_2024}.

The precise in-plane orientation of the ordered Mn moments in bulk crystals has nevertheless remained under debate.
Below $T_\mathrm{N}$, threefold rotation symmetry permits three in-plane N\'eel-vector orientations separated by $120^\circ$; in addition, time-reversal symmetry operation generates an antiphase counterpart for each in-plane orientation, giving rise to six magnetic domain states in total~\cite{takegami_circular_2025,liu_altermagnetic_order_2026,huynh_intrinsic_2026} (Fig.~\ref{fig:fig1}\textbf{(e)}).
While the presence of these orientational domains obscures the precise determination of the ordered moment direction in conventional diffraction experiment, a balanced antiphase domain population could cancel out experimental signatures in time-reversal-odd transport and optical responses, and in polarised-neutron detection of chiral magnon splitting~\cite{mazin_altermagnetism_2023,mcclarty_observing_2025,liu_altermagnetic_order_2026}.
The energy-split chiral magnon bands could be observed even without achieving a single domain state~\cite{liu_chiral_2024}, but its chiral polarization requires the presence of a finite sample-averaged N\'eel vector.
Such dynamical magnon chirality, a defining feature for the emergence of altermagnetism, was first verified in domain-biased $d$-wave altermagnet MnF$_2$~\cite{faure_altermagnetism_2025} and was subsequently observed in $g$-wave altermagnet $\alpha$-MnTe by polarized inelastic neutron scattering after cooling under millitesla-scale field~\cite{liu_switchable_2026}.
By contrast, circular dichroism signals in resonant inelastic x-ray scattering primarily reflect static domain imbalance rather than dynamical magnon chirality, as recently demonstrated in both CrSb and $\alpha$-MnTe~\cite{biniskos_systematic_2025,takegami_circular_2025}.

Several approaches to achieve domain control in $\alpha$-MnTe have been reported.
Kriegner \textit{et al.} combined neutron diffraction, magnetotransport, and magnetometry on epitaxial thin films to determine the basal-plane easy axis as $[1, \bar{1}, 0]$~\cite{kriegner_magnetic_2017}, although the thin films contain a range of epitaxial strain states, the biaxial strain mainly shifts the spin-reorientation threshold from approximately 2~T on InP to 0.5~T on SrF$_2$, rather than acting as the detwinning stimulus itself.
In bulk crystals, it has been found that stain combined with field cooling selects time-reversed domains through the piezomagnetic response~\cite{aoyama_piezomagnetic_2024}, and that a uniaxial compression could lead to the detwinning of the orientational domains thus enabling the strain control of the anomalous Hall effect~\cite{liu_strain-tunable_2026}.
Spatially resolved magneto-optical measurements instead find that strain can rotate the local N\'eel vector continuously~\cite{liebmanpelaez_strain_2026}.
Chemical substitution provides a different control route: less than 1\% Li doped in MnTe rotates the moments from the basal plane toward the $c$ axis, removing the threefold in-plane degeneracy~\cite{moseley_doping_2022,yumnam_magnon_2024,zhang_li-doped_2025}.

Recent experiments have substantially broadened the range of domain-sensitive probes.
Polarized neutron diffraction established millitesla-scale switching of antiphase domains by $c$-axis field cooling and partial orientational selection by oblique-field cooling, with a net N\'eel-vector magnitude of about 12\%~\cite{liu_altermagnetic_order_2026}, the same antiphase control was used to reverse the measured magnon chirality~\cite{liu_switchable_2026}.
Scanning Kerr microscopy has directly imaged macroscopic altermagnetic domains in bulk crystals and demonstrated their controllability and stability against magnetic and thermal perturbations~\cite{watanabe_magneto-optical_2026}.
Nitrogen-vacancy magnetometry has directly imaged the surface magnetization of MnTe films and demonstrated field control of the domain population~\cite{zhou_surface_2026}.
Complementary NMR, magnetostriction and inelastic-neutron measurements show that in-plane fields depopulate unfavorable orientations and continuously rotate the surviving N\'eel vectors toward a field-perpendicular configuration~\cite{duan_magnetostriction_2026,weber_magnetic_2026,huynh_intrinsic_2026}.
The principal experimental control strategies are summarised and compared in Table~\ref{tab:comparison}.

In this work, we present a detailed single-crystal neutron diffraction study of bulk $\alpha$-MnTe using an in-plane magnetic field applied along the crystallographic $[\bar{1},1,0]$ direction.
We demonstrate that a moderate in-plane field of $\sim 2$~T induces an efficient detwinning of the orientational domains, achieving a domain population imbalance of $r_A:r_B:r_C \approx 0.97:0.03(5):0$, i.e., $\sim 97\%$ detwinning efficiency.
Through the magnetic structure refinement based on 61 reflections collected at 2~K under the detwinned condition, we determine that the ordered Mn moments are aligned along the $[1,1,0]$ direction.
This moment orientation under in-plane field lies $30^\circ$ away from the $[1, \bar{1}, 0]$ direction found in strain-based detwinning process~\cite{liu_strain-tunable_2026} and in epitaxial thin films~\cite{kriegner_magnetic_2017}. Furthermore, an abnormal behaviour in the temperature dependence of the selected reflections is observed in 60--88~K, while an extrinsic cause due to the minority impurity phase can be ruled out, its microscopic origin, discussed within the context of possible intrinsic piezomagnetic or magnetoelastic effects, remains open.
Our approach provides a practical, non-contact pathway to prepare orientational-domain-free bulk $\alpha$-MnTe. Combined with antiphase-domain selection, this can lead to an efficient preparation of truly single-domain samples, thus enabling future studies of altermagnetic phenomena with a wide range of experimental techniques.

\section{Experimental details}

\begin{figure*}[ht!]
    \includegraphics[width=\linewidth]{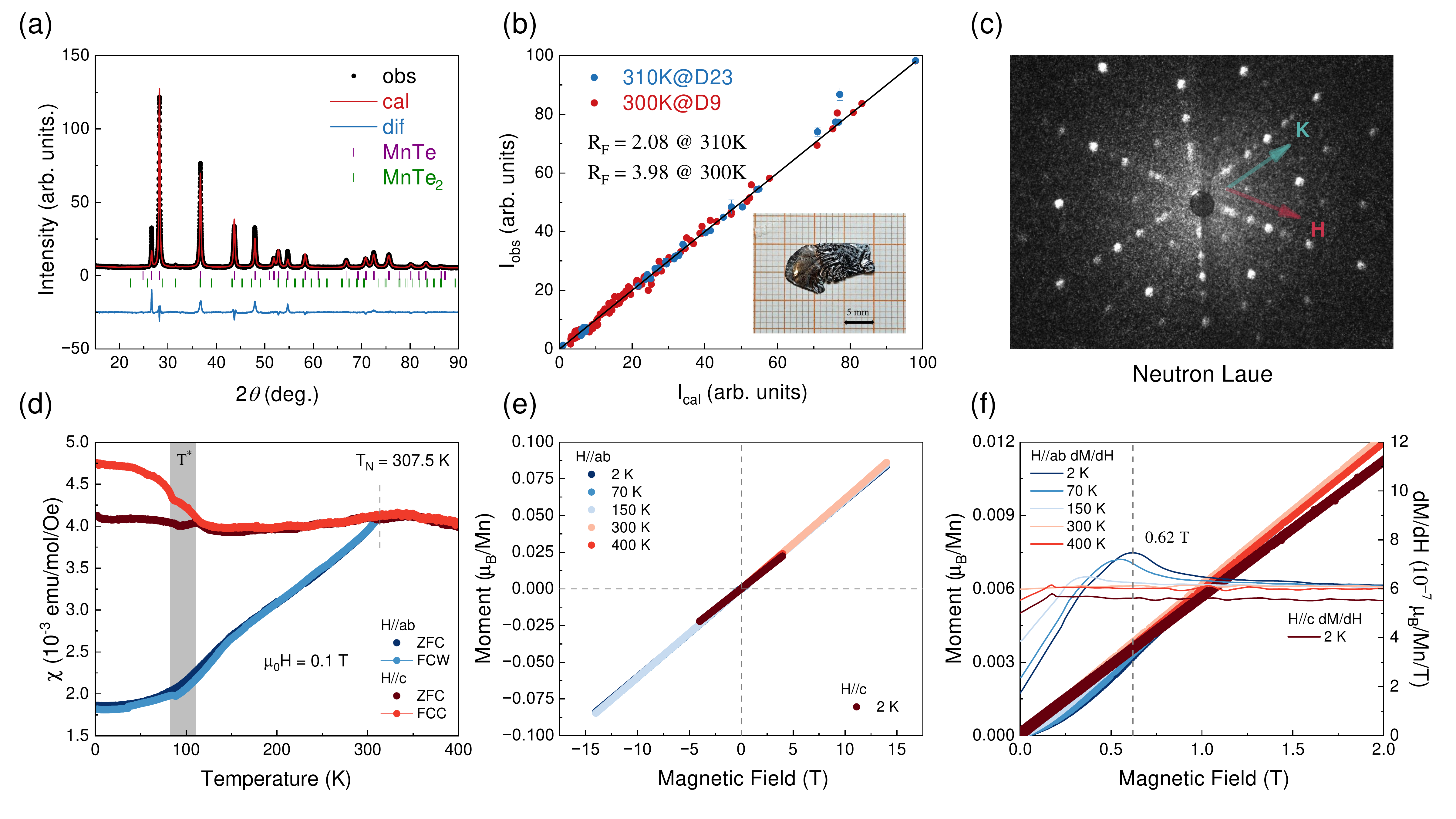}
    \caption{
    \textbf{(a)} Powder X-ray diffraction pattern and Rietveld refinement; tick marks indicate the $\alpha$-MnTe and MnTe$_2$ ($<0.05\%$) phases.
    \textbf{(b)} Single-crystal neutron diffraction refinement ($I_\mathrm{obs}$ vs.\ $I_\mathrm{cal}$) collected at D23 (310~K, $R_F = 2.08$) and D9 (300~K, $R_F = 3.98$); the inset shows a photograph of a typical plate-like crystal.
    \textbf{(c)} Single-crystal neutron Laue pattern collected on OrientExpress at ILL.
    \textbf{(d)} Temperature dependence of the magnetic susceptibility at $\mu_0H=0.1$~T for $\mathbf{H}\parallel c$ and $\mathbf{H}\parallel ab$ after zero-field cooling (ZFC) and during field-cooled cooling or warming (FCC or FCW); the grey region marks $T^*=82.5$--106.5~K and the dashed line $T_\mathrm{N}=307.5$~K.
    \textbf{(e)} Full view of the magnetization as a function of field for $\mathbf{H}\parallel ab$ and $\mathbf{H}\parallel c$ at various temperatures.
    \textbf{(f)} Low-field magnetization for $\mathbf{H}\parallel ab$ and $\mathbf{H}\parallel c$ at various temperatures, together with the derivative $dM/dH$ (right axis); the dashed line marks the $dM/dH$ maximum at 0.62~T at 2~K.}
    \label{fig:fig2}
\end{figure*}

Single-crystal $\alpha$-MnTe was grown using the self-flux method.
The starting molar ratio of Mn:Te was $29.6:70.4$.
A total of 16~g of high-purity elements was weighed and mixed in a glove box, loaded into an aluminium crucible, evacuated and sealed in a silica ampoule.
The ampoule was placed in a vertical furnace and heated to $950~^{\circ}$C at a rate of $100~^{\circ}$C/h, held for 10~h, and subsequently cooled to $780~^{\circ}$C over 300~h.
At this temperature, the flux was removed by centrifugation.
The resulting crystals are plate-like with a typical dimension of $15~\mathrm{mm} \times 9~\mathrm{mm} \times 0.8~\mathrm{mm}$ (see Fig.~\ref{fig:fig2}\textbf{(b)} inset).

Powder X-ray diffraction (XRD) measurements were performed using a Bruker D2 diffractometer.
Magnetic susceptibility and magnetization were measured using Quantum Design physical property measurement system (PPMS).
The sample mass used for magnetic measurements is around 52~mg.

\begin{figure*}[p]
    \includegraphics[width=\linewidth]{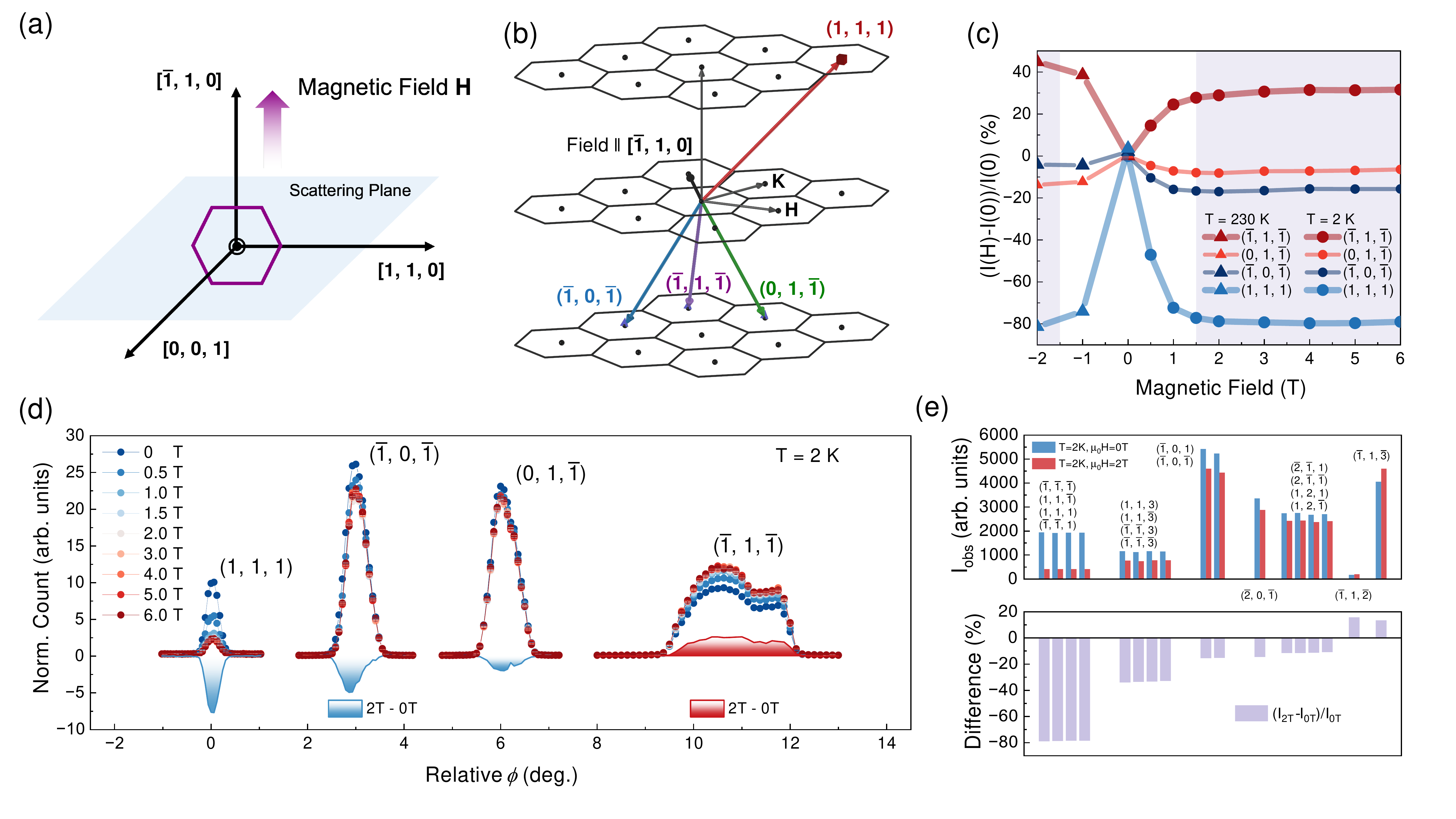}
    \caption{
    \textbf{(a)} Schematic of the scattering plane and the applied magnetic field direction in the neutron diffraction experiment.
    \textbf{(b)} 3D reciprocal space map of MnTe with the labeled reflections tracked in this study.
    \textbf{(c)} Relative change of the integrated intensities, $(I(H)-I(0))/I(0)$, of the $(1,1,1)$, $(0,1,\bar{1})$, $(\bar{1},0,\bar{1})$, and $(\bar{1},1,\bar{1})$ reflections with $\mathbf{H}\parallel[\bar{1},1,0]$.
    The positive-field branch (0 to 6~T) was measured at $T = 2$~K; the negative branch shows the field-reversal protocol ($+6 \to 0 \to -2$~T) at $T = 230$~K, demonstrating full recovery of the initial intensities at 0~T and re-saturation by $-2$~T (the positive-field branch at 230~K, omitted for clarity, closely follows the 2~K behavior).
    The $(\bar{1},1,\bar{1})$ reflection increases while the others decrease, in agreement with the refinement results.
    \textbf{(d)} Raw $\phi$ scans at $T=2$~K for $\mu_0H=0$, 0.5, 1.0, 1.5, 2.0, 3.0, 4.0, 5.0, and 6.0~T; the 2~T minus 0~T curves emphasize the field-induced changes.
    \textbf{(e)} Integrated intensities at 0~T and 2~T (top) and their relative change (bottom) for the 17 most strongly affected reflections; equivalent reflections are grouped and sorted by the change.}
    \label{fig:fig3}
\end{figure*}

Single-crystal neutron diffraction measurements were carried out on the D23 and D9 instruments at the Institut Laue--Langevin (ILL).
The neutron wavelengths used are 1.283~\AA{} (D23) and 0.834~\AA{} (D9), respectively.
The sample is oriented in the $(H,H,L)$ scattering plane (Fig.~\ref{fig:fig3}\textbf{(a)}), with the magnetic field along $[\bar{1},1,0]$.
With the advantage of a lifting-counter detector on both D23 and D9, we were able to access reflections out of the scattering plane.
Diffraction data were collected over a range of temperatures and fields to determine the magnetic structure and track its field-induced evolution.

Magnetic structure refinement was performed with \textsc{Mag2Pol}~\cite{qureshi_mag2pol_2019} using nuclear and magnetic structure factors for each reflection.
The three orientational domain populations are constrained to be non-negative and to sum to unity.
They are generated by the following symmetry operators: $(x,\, y,\, z)$, $(-y,\, x-y,\, z)$ and $(y-x,\, -x,\, z)$.

\section{Structure and magnetic properties}

The crystal and magnetic configurations of $\alpha$-MnTe are summarized in Fig.~\ref{fig:fig1}.
The powder X-ray diffraction pattern and its quantitative multiphase refinement (Fig.~\ref{fig:fig2}\textbf{(a)}) confirm the hexagonal $P6_3/mmc$ structure and place the MnTe$_2$ impurity fraction about 0.5 wt\%.
Mn occupies the $2a$ Wyckoff site and Te the $2c$ site, producing triangular Mn layers in the $ab$ plane (Fig.~\ref{fig:fig1}\textbf{(a)}).
Single-crystal neutron refinements at 310~K on D23 and 300~K on D9 (Fig.~\ref{fig:fig2}\textbf{(b)}), together with the Laue pattern (Fig.~\ref{fig:fig2}\textbf{(c)}), demonstrate the high crystallographic quality of the sample. Recent electron microscopy, second-harmonic-generation polarimetry, and Raman measurements have nevertheless identified a noncentrosymmetric structural distortion within local structural domains~\cite{wu_noncentrosymmetric_2026}.
The lattice parameters refined from the 300~K D23 data are $a=4.1477(13)$~\AA{} and $c=6.7044(26)$~\AA{}.

Below $T_\mathrm{N}$, the Mn moments are ferromagnetically aligned within each $ab$ layer and antiferromagnetically stacked along $c$ (Fig.~\ref{fig:fig1}\textbf{(a,b)}).
The opposite-spin layers are related neither by translation nor by inversion, but by the spin-space-group operation $[C_2\parallel C_6t_{1/2}]$, which combines a twofold spin rotation with a sixfold screw operation (Fig.~\ref{fig:fig1}\textbf{(b)}).
Consequently, $\mathcal{P}\mathcal{T}$ symmetry is absent and momentum-dependent spin splitting is allowed despite the compensated moment~\cite{smejkal_beyond_2022,chen_unconventional_2025}.

Different in-plane moment directions correspond to distinct magnetic symmetries and macroscopic responses.
Using the Bilbao Crystallographic Server convention, the $[1,1,0]$ configuration is associated with magnetic space group $Cmcm$ (No.~63.457; point group $mmm$), whereas $[\bar{1},1,0]$ is associated with $Cm'c'm$ (No.~63.462; point group $m'm'm$) (Fig.~\ref{fig:fig1}\textbf{(c,d)})~\cite{aroyo_bilbao_2006,lovesey_templates_2023}.
The latter permits an antisymmetric $\sigma_{xy}$ response, the anomalous Hall and magneto-optical Kerr effects, and a weak $c$-axis ferromagnetic moment~\cite{gonzalez_betancourt_spontaneous_2023,kluczyk_coexistence_2024,yang_ultrafast_2026}.
The $Cmcm$ configuration forbids these time-reversal-odd responses but permits the time-even Voigt effect~\cite{lovesey_templates_2023,yang_ultrafast_2026}. The giant spontaneous Kerr rotation observed in bulk crystals is, however, absent in stoichiometric insulating films, indicating that carrier self-doping underlies the macroscopic magneto-optical response~\cite{yang_kerr_2026}.
Consistently, the leading angular dependence is $\sigma_{xy}\propto\sin 3\alpha$, where $\alpha$ is measured from a nearest-neighbor Mn--Mn bond: it vanishes for the $[1,1,0]$ family and is maximal for the $[\bar{1},1,0]$ family~\cite{mazin_altermagnetism_2023}.

The three orientational variants and their time-reversed partners are illustrated in Fig.~\ref{fig:fig1}\textbf{(e)}.
Selecting one orientation while retaining both antiphase partners produces the state in Fig.~\ref{fig:fig1}\textbf{(f)}; selecting one antiphase sign while retaining multiple orientations produces Fig.~\ref{fig:fig1}\textbf{(g)}; and selecting both yields the fully single-domain state in Fig.~\ref{fig:fig1}\textbf{(h)}.
An in-plane field lowers the free energy of configurations whose N\'eel vector approaches the field-perpendicular direction, thereby lifting the orientational degeneracy through a coupling that is even in the field~\cite{huang_field-selection_2026}.

Figure~\ref{fig:fig2}\textbf{(d)} shows the magnetic susceptibility for fields in the $ab$ plane and along $c$, yielding $T_\mathrm{N}=307.5$~K.
For $\mathbf{H}\parallel c$, the ZFC and FCC curves separate over approximately 82.5--106.5~K.
A related susceptibility rise near 83~K and resistivity upturn below 100~K were reported in polycrystalline MnTe without an accompanying structural transition~\cite{efrem_dsa_low-temperature_2005}.
Angle-dependent magnetization has more recently identified spontaneous magnetization below approximately 81~K together with a $90^\circ$ change of the easy-magnetization axis~\cite{orlova_magnetization_2025}.
We compare these observations with our diffraction data below.

The magnetization remains far from saturation up to 14~T for $\mathbf{H}\parallel ab$ (Fig.~\ref{fig:fig2}\textbf{(e)}), as expected for an antiferromagnet.
At fields below about 1~T, however, the in-plane response is nonlinear at 2, 70, and 150~K.
The corresponding $dM/dH$ curves (Fig.~\ref{fig:fig2}\textbf{(f)}) have maxima that shift to lower field with increasing temperature; at 2~K the maximum occurs at 0.62~T and the response approaches its high field slope by approximately 1.5~T.
This behavior is consistent with field-driven reorientation and redistribution of the in-plane magnetic configurations, as examined directly by diffraction below~\cite{rossler_lowfield_2025}.

\section{Neutron diffraction under in-plane magnetic field}
\subsection{Field-induced detwinning of orientational domains}

Comprehensive neutron diffraction experiments under an applied in-plane magnetic field were carried out. Figure~\ref{fig:fig3}\textbf{(a)} shows the scattering plane and the applied-field direction, and Fig.~\ref{fig:fig3}\textbf{(b)} marks the reflections tracked in reciprocal space.

To follow the field-induced orientational response, we measured the $(1,1,1)$, $(0,1,\bar{1})$, $(\bar{1},0,\bar{1})$, and $(\bar{1},1,\bar{1})$ magnetic intensities as a function of an in-plane field along $[\bar{1},1,0]$ at 2~K (Fig.~\ref{fig:fig3}\textbf{(c)}).
The corresponding rocking scans are shown in Fig.~\ref{fig:fig3}\textbf{(d)}.

All four intensities evolve smoothly and saturate by approximately 2~T.
The $(1,1,1)$ reflection decreases by about 80\%, while $(\bar{1},1,\bar{1})$ increases by 32\%; the changes of $(0,1,\bar{1})$ and $(\bar{1},0,\bar{1})$ are approximately $-6\%$ and $-16\%$, respectively.
This pattern is consistent with a common moment direction $[1,1,0]$, perpendicular to the applied field.
In an easy-plane antiferromagnet, the field-dependent susceptibility lowers the free energy as the N\'eel vector approaches the field-perpendicular direction, so that the response is even in the field and saturates once the N\'eel vector is fully aligned perpendicular to the field.

The reversal protocol at 230~K provides an independent test of the field symmetry (Fig.~\ref{fig:fig3}\textbf{(c)}).
After the initial sweep from 0 to $+6$~T, the field was reduced through zero to $-2$~T, the omitted positive-field branch at 230~K closely follows the 2~K curve.
The intensities return to their initial values at zero field and reach the same saturated configuration by $-2$~T.
This reversible, field-even response is expected for orientational alignment~\cite{huang_field-selection_2026}.
Likewise, the absence of further intensity changes from 2 to 6~T shows that no additional field-induced reorganization occurs above the alignment field; it does not exclude the spin-reorientation process responsible for the low-field evolution itself~\cite{kriegner_magnetic_2017,duan_magnetostriction_2026,weber_magnetic_2026,huynh_intrinsic_2026}.

The field-induced changes extend well beyond the four reflections discussed above.
Fig.~\ref{fig:fig3}\textbf{(e)} compares the 0 and 2~T intensities of the 17 most strongly affected reflections, grouped by crystallographic equivalence.
The four equivalents of $(1,1,1)$ each decrease by approximately 80\%, while reflections in the $(\bar{1},1,\bar{l})$ family increase.
The internally consistent changes support the common $[1,1,0]$ moment orientation.

\subsection{Magnetic structure refinement}

\begin{figure}[ht!]
    \includegraphics[width=\linewidth]{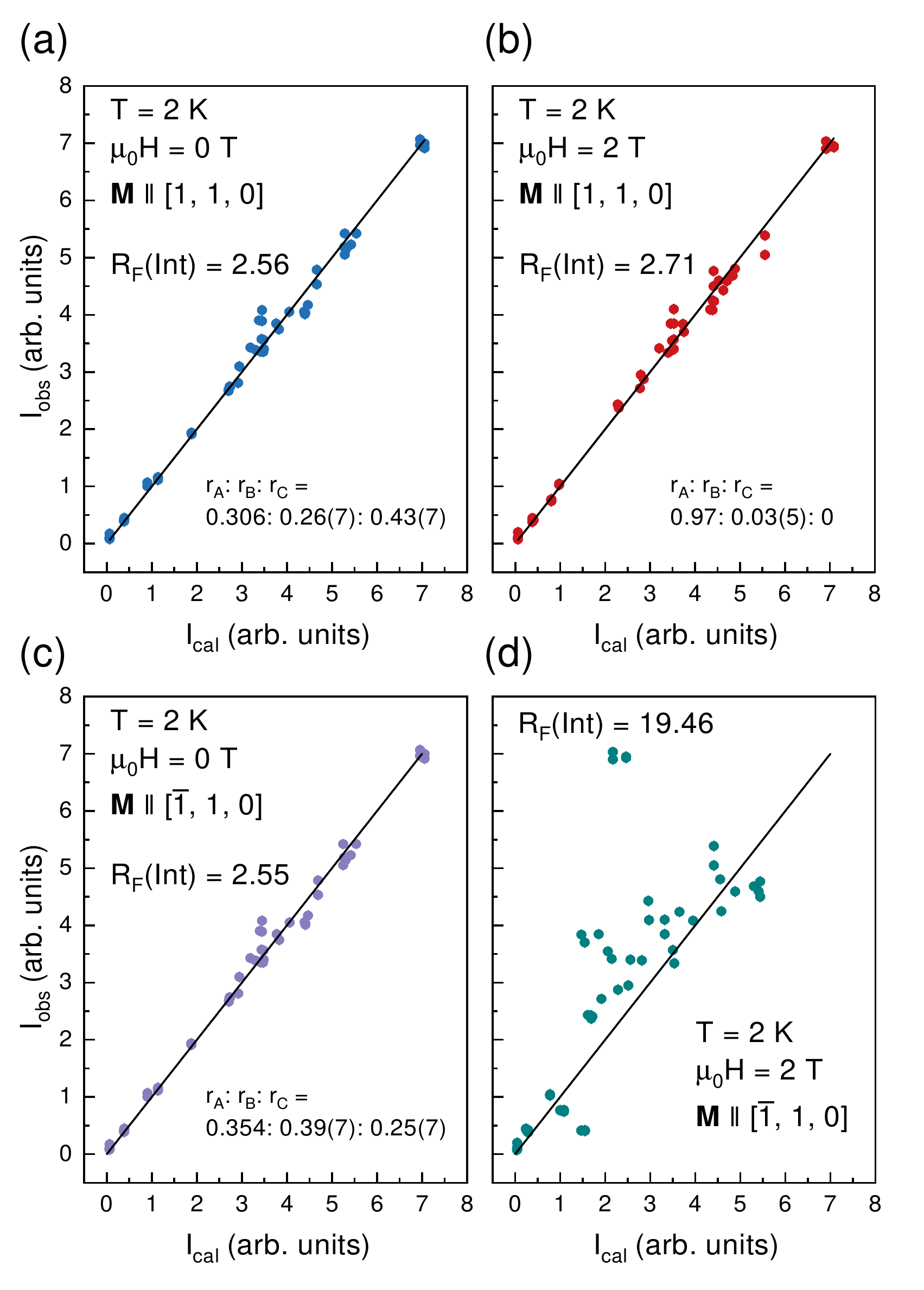}
    \caption{
    \textbf{(a)} $I_\mathrm{obs}$ versus $I_\mathrm{cal}$ at 2~K and 0~T, assuming $\mathbf{M}\parallel[1,1,0]$, with $R_F(\mathrm{Int})=2.56$ and $r_A:r_B:r_C=0.306:0.26(7):0.43(7)$, statistically compatible with an approximately balanced orientational population.
    \textbf{(b)} Constrained three-orientation refinement at 2~K and 2~T, yielding $R_F(\mathrm{Int})=2.71$, $\mathbf{M}\parallel[1,1,0]$, $M=4.62(8)~\mu_\mathrm{B}$, and $r_A:r_B:r_C=0.97:0.03(5):0$; $r_C$ reaches its nonnegative lower bound.
    \textbf{(c)} $I_\mathrm{obs}$ versus $I_\mathrm{cal}$ at 2~K and 0~T, assuming $\mathbf{M}\parallel[\bar{1},1,0]$, with $R_F(\mathrm{Int})=2.55$ and $r_A:r_B:r_C=0.354:0.39(7):0.25(7)$, statistically compatible with an approximately balanced orientational population.
    \textbf{(d)} Alternative refinement with $\mathbf{M}\parallel[\bar{1},1,0]$ at 2~K and 2~T, showing substantially poorer agreement, $R_F(\mathrm{Int})=19.46$.}
    \label{fig:fig4}
\end{figure}

To determine the moment direction in the aligned state, we collected 61 nuclear and magnetic reflections at 2~K in a 2~T field along $[\bar{1},1,0]$.
Within the constrained three-orientation model, the best fit has $\mathbf{M}\parallel[1,1,0]$, $M=4.62(8)~\mu_\mathrm{B}$ per Mn, and $R_F(\mathrm{Int})=2.71$ (Fig.~\ref{fig:fig4}\textbf{(b)}).
The ordered moment is slightly below the spin-only value of $5~\mu_\mathrm{B}$ for high-spin Mn$^{2+}$ but lies within the range reported by earlier neutron studies~\cite{kunitomi_neutron_1964,kriegner_magnetic_2017}.
Ignoring the explicit symmetry reduction produced by the applied field, the magnetic configuration associated with this moment direction is $Cmcm$ (No.~63.457)~\cite{aroyo_bilbao_2006,lovesey_templates_2023}.
The refined orientational populations are $r_A:r_B:r_C=0.97:0.03(5):0$, with $r_C$ at the nonnegative lower bound.
The minority population is therefore statistically compatible with zero and the aligned state is best described as orientational-domain-free.

\begin{table}[ht!]
\caption{Structural and magnetic parameters refined with \textsc{Mag2Pol} at $\mu_0H=2$~T along $[\bar{1},1,0]$. The space group is $P6_3/mmc$ (No.~194), and the moment direction is $[1,1,0]$.}
\label{refinement_table1}
\begin{ruledtabular}
\begin{tabular}{lcc}
Parameter & 2~K & 200~K \\ \hline
$a=b$ (\AA) & 4.1253(11) & 4.1369(10) \\
$c$ (\AA) & 6.6539(21) & 6.6799(19) \\
$r_A:r_B:r_C$ & $0.97:0.03(5):0$ & $0.954:0.046:0$ \\
$M_\mathrm{Mn}$ ($\mu_\mathrm{B}$) & 4.62(8) & 3.94(6) \\
$R_F(\mathrm{Int})$ & 2.71 & 2.59 \\
\end{tabular}
\end{ruledtabular}
\end{table}

The refinements at 2 and 200~K used the same crystallographic model.
Mn and Te positions were fixed by $P6_3/mmc$ symmetry at $(0,0,0)$ and $(1/3,2/3,1/4)$, respectively; occupancies were fixed to unity and isotropic displacement parameters to zero because trial refinements produced negligible, unstable values.
The temperature-dependent parameters are collected in Table~\ref{refinement_table1}.
For reference, the corresponding lattice parameters at 300~K are $a=4.1477(13)$~\AA{} and $c=6.7044(26)$~\AA{}.

At zero field, $r_A:r_B:r_C=0.306:0.26(7):0.43(7)$ assuming $\mathbf{M}\parallel[1,1,0]$ (Fig.~\ref{fig:fig4}\textbf{(a)}), and $r_A:r_B:r_C=0.354:0.39(7):0.25(7)$ assuming $\mathbf{M}\parallel[\bar{1},1,0]$ (Fig.~\ref{fig:fig4}\textbf{(c)}). Both solutions are compatible with nearly equal populations within the quoted uncertainties, so the initial state is approximately balanced and the zero-field moment direction cannot be distinguished, with both models giving $R_F(\mathrm{Int})$ values near $2.55\%$.

The $[1,1,0]$ direction belongs to the nearest-neighbor (NN) Mn--Mn bond family.
The easy axes reported for the epitaxial films and the strain-controlled bulk study belong instead to the NNN family, represented here by $[\bar{1},1,0]$. The two symmetry families are rotated by $30^\circ$ relative to one another, the representatives $[1,1,0]$ and $[\bar{1},1,0]$ are themselves $90^\circ$ apart~\cite{kriegner_magnetic_2017,liu_strain-tunable_2026}.
Fixing the moment along $[\bar{1},1,0]$ under the same field conditions gives $R_F(\mathrm{Int})=19.46\%$ and the pronounced misfit in Fig.~\ref{fig:fig4}\textbf{(d)}, ruling out that field-parallel configuration.

Table~\ref{table:magnetic_space_group} summarizes the magnetic space groups associated with the two moment configurations. A critical difference is that the $Cm'c'm$ configuration permits a weak $c$-axis ferromagnetic moment, whereas the $Cmcm$ configuration does not, and that the $(h,h,0)$ and $(h,-h,0)$ reflections are magnetically forbidden for $Cmcm$; the temperature dependence of the $(1,1,0)$ reflection is discussed in the next section.
Polarized neutron diffraction directly accesses the antiphase domain imbalance and has shown that $c$-axis field cooling reverses it, while oblique-field cooling also modifies the orientational population~\cite{liu_altermagnetic_order_2026}.
Combining these two controls may therefore realize a fully single-domain state, which would be of interest for future studies of phenomena that are otherwise obscured by domain averaging.

\begin{table*}[ht!]
\caption{Magnetic space groups associated with the two in-plane moment configurations, obtained with the Bilbao Crystallographic Server~\cite{aroyo_bilbao_2006}. Absences refer to magnetic contributions in the parent hexagonal indexing.}
\label{table:magnetic_space_group}
\begin{ruledtabular}
\begin{tabular}{llll}
Magnetic space group & BNS No. & Mn positions and moments & Magnetic absences \\ \hline
$Cmcm$ & 63.457 & \tcell{7.0cm}{$(0,0,0\mid m_a,m_a,0)$; $(0,0,1/2\mid-m_a,-m_a,0)$} & \tcell{3.0cm}{$(00l)$, $l=2n$; \\ $(hh0)$; $(h\bar{h}0)$} \\
$Cm'c'm$ & 63.462 & \tcell{7.0cm}{$(0,0,0\mid m_a,-m_a,m_c)$; $(0,0,1/2\mid-m_a,m_a,m_c)$} & \tcell{3.0cm}{$(00l)$, $l=2n$} \\
\end{tabular}
\end{ruledtabular}
\end{table*}

\subsection{Temperature dependence of the selected reflections}

\begin{figure*}[t!]
    \includegraphics[width=\linewidth]{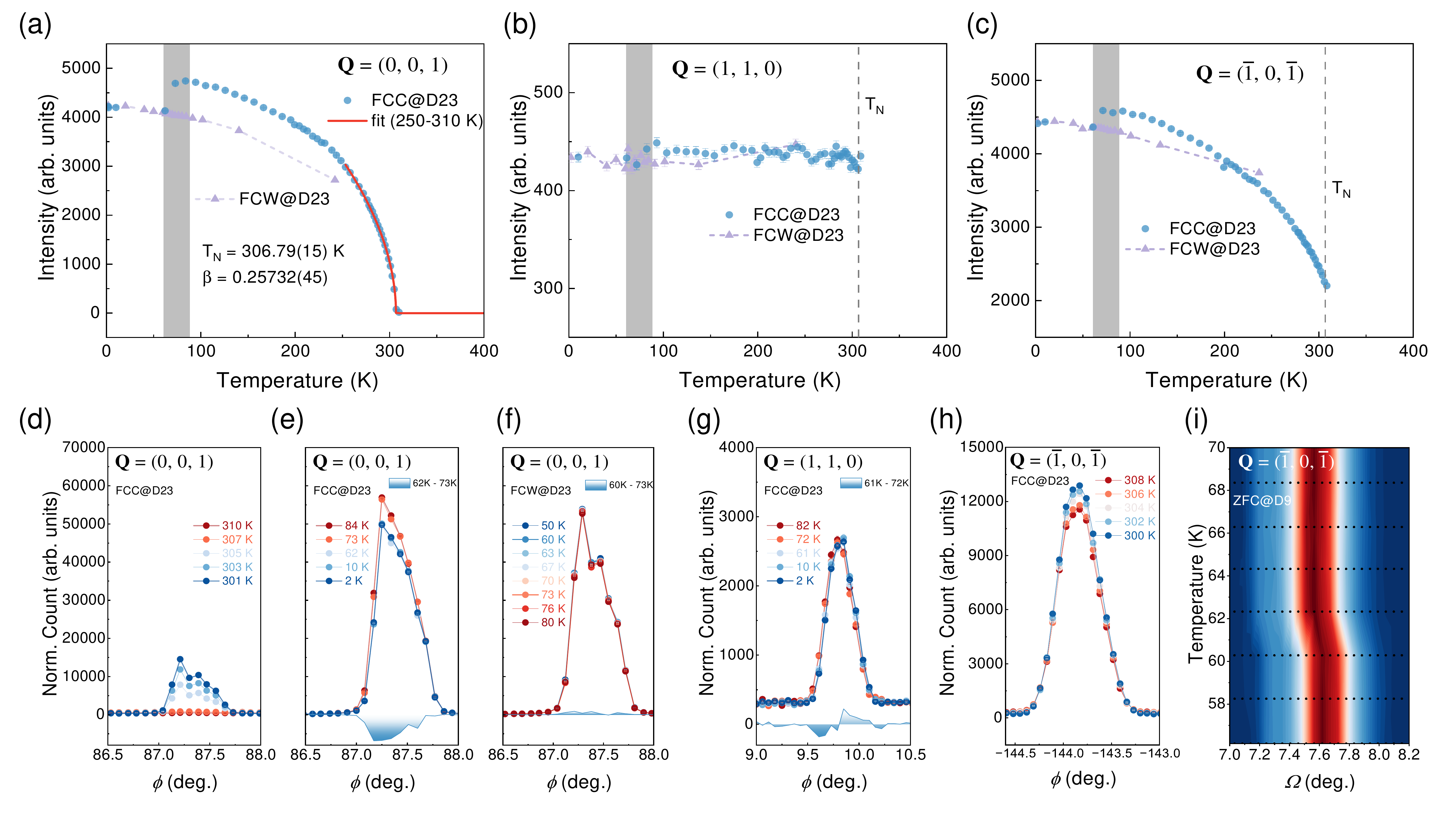}
    \caption{
    \textbf{(a)} Integrated intensity of the $(0,0,1)$ reflection measured during field-cooled cooling (FCC) on D23, fitted over 250--310~K to $I=I_0[(T_\mathrm{N}-T)/T_\mathrm{N}]^{2\beta}$ with $T_\mathrm{N}=306.79(15)$~K and $\beta=0.25732(45)$; the grey region marks 60.5--88.5~K.
    \textbf{(b)} Integrated intensity of the $(1,1,0)$ reflection; no resolvable magnetic contribution or anomaly is observed.
    \textbf{(c)} Integrated intensity of the $(\bar{1},0,\bar{1})$ reflection ;
    \textbf{(d)} Raw $\phi$ rocking scans of $(0,0,1)$ across $T_\mathrm{N}$ (310--301~K, FCC@D23).
    \textbf{(e)} Raw $\phi$ scans of $(0,0,1)$ at selected temperatures from 84 to 2~K (FCC@D23).
    \textbf{(f)} Raw $\phi$ scans of $(0,0,1)$ from 50 to 80~K during field-cooled warming (FCW@D23).
    \textbf{(g)} Raw $\phi$ scans of $(1,1,0)$ at selected temperatures from 82 to 2~K (FCC@D23).
    \textbf{(h)} Raw $\phi$ rocking scans of $(\bar{1},0,\bar{1})$ across $T_\mathrm{N}$ (308--300~K, FCC@D23).
    \textbf{(i)} Temperature-dependent $\Omega$-scan intensity map of $(\bar{1},0,\bar{1})$ from 70 to 58~K with zero-field cooling (ZFC@D9).}
    \label{fig:fig5}
\end{figure*}

The $(0,0,1)$ reflection is purely magnetic: the nuclear structure factor of $\alpha$-MnTe contributes only to the reflections with even $l$; and this scattering vector, being parallel to $c$, probes only the ordered in-plane moment component.
Its onset below $T_\mathrm{N}$ is evident both in the integrated intensity (Fig.~\ref{fig:fig5}\textbf{(a)}) and in the raw rocking scans (Fig.~\ref{fig:fig5}\textbf{(d)}).
A power-law fit over 250--310~K gives $T_\mathrm{N}=306.79(15)$~K, consistent with the transition temperature seen in the susceptibility, and an fitted critical exponent $\beta=0.25732(45)$ that is numerically close to the mean-field tricritical value $\beta=0.25$~\cite{riedel_tricritical_1972,wegner_logarithmic_1973} but differs substantially from the previous powder neutron value of $0.317(20)$~\cite{yumnam_magnon_2024}.

The $(1,1,0)$ position contains nuclear scattering but would acquire a magnetic contribution from a $c$-axis component.
No such temperature-dependent contribution is resolved (Fig.~\ref{fig:fig5}\textbf{(b)}), suggesting that the ordered moment is strictly aligned within the basal plane within the sensitivity of this measurement.
Magnetic scattering at $(1,1,0)$ is forbidden for $Cmcm$ but allowed for $Cm'c'm$, as summarized in Table~\ref{table:magnetic_space_group}: for the latter, the symmetry-permitted $c$-axis moment contributes to $(1,1,0)$, although symmetry does not require that moment to be finite~\cite{aroyo_bilbao_2006,lovesey_templates_2023}.
Thus, the result is consistent with, but does not by itself establish, the $Cmcm$ configuration refined above.

The FCC data show a history-dependent decrease of the $(0,0,1)$ intensity between approximately 60.5 and 88.5~K (Fig.~\ref{fig:fig5}\textbf{(a)}).
Here FCC denotes cooling from 310~K in $\mu_0H=2$~T while collecting data, FCW denotes measurement on warming after the same field-cooling protocol, and ZFC denotes zero-field cooling.
The abrupt decrease is visible in the FCC rocking scans (Fig.~\ref{fig:fig5}\textbf{(e)}) but is not reproduced in FCW (Fig.~\ref{fig:fig5}\textbf{(f)}), thus arguing against a reversible equilibrium phase transition.
Because $(1,1,0)$ does not gain magnetic intensity over the same temperature range (Figs.~\ref{fig:fig5}\textbf{(b)} and \ref{fig:fig5}\textbf{(g)}), the decrease is not accompanied by a resolvable rotation of the moments out of the basal plane.
A similar history-dependent decrease is observed in the $(\bar{1},0,\bar{1})$ reflection (Fig.~\ref{fig:fig5}\textbf{(c)}), which mixes nuclear and magnetic scattering (Fig.~\ref{fig:fig5}\textbf{(h)}).

The temperature window overlaps approximately with the onset of the $c$-axis susceptibility irreversibility and the magnetization anomaly at around 81--83~K~\cite{efrem_dsa_low-temperature_2005,orlova_magnetization_2025}.
Efrem D'Sa \textit{et al.} found no structural transition or out-of-plane spin tilt, but reported an increase of $c/a$ below 75~K and attributed the associated susceptibility and transport anomalies to magnetoelastic coupling~\cite{efrem_dsa_low-temperature_2005}. The non-centrosymmetric structural distortion within local domains reported recently~\cite{wu_noncentrosymmetric_2026} indicates, however, that subtle structural effects in MnTe deserve further attention.
Our $\Omega$-scan map of $(\bar{1},0,\bar{1})$ measured in ZFC shows a small shift of the rocking-curve center over the same temperature region (Fig.~\ref{fig:fig5}\textbf{(i)}).
Because a rocking-angle shift can also arise from sample reorientation or thermal motion, it is compatible with but is not direct evidence for a lattice parameter anomaly.
\begin{figure}[t!]
    \includegraphics[width=\linewidth]{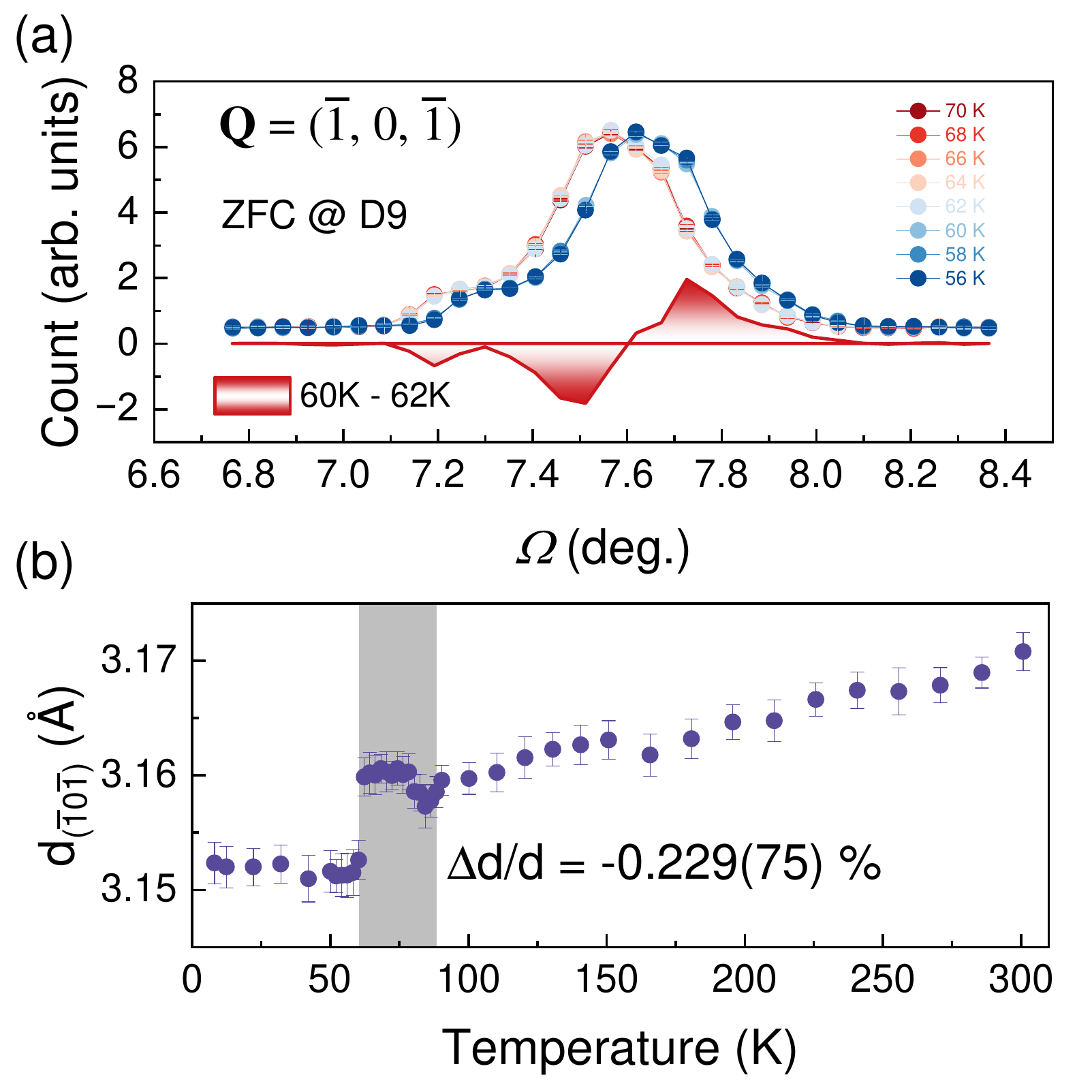}
    \caption{
    \textbf{(a)} Raw $\Omega$ scans recorded on D9 during zero-field cooling at 70, 68, 66, 64, 62, 60, 58 and 56~K; the red curve is the difference between the 60 and 62~K scans.
    \textbf{(b)} $d$-spacing of $(\bar{1},0,\bar{1})$ versus temperature, derived from the detector-spot centroid and the recorded arm angle; the grey band marks 60--80~K and the annotation gives the pairwise step $\Delta d/d=-0.229(75)\%$.}
    \label{fig:fig6}
\end{figure}

To settle this point, we quantified the peak position in the ZFC data taken at D9. For each $(\bar{1},0,\bar{1})$ rocking scan, the spot centroid on the position-sensitive detector together with the recorded detector-arm angle yields an absolute scattering angle and hence a $d$-spacing (Fig.~\ref{fig:fig6}). Between the consecutive scans at 62.33 and 60.28~K, $d_{(\bar{1},0,\bar{1})}$ drops abruptly by $\Delta d/d=-0.229(75)\%$, while the integrated intensity and the peak width remain continuous.

Such an event could be a natural consequence of the intrinsic piezomagnetic effect of the altermagnetic state~\cite{aoyama_piezomagnetic_2024}. The piezomagnetic tensor, which couples the magnetic order linearly to the strain, is nonzero by symmetry and reverses sign between two time-reversal domain states, so that the equilibrium lattice spacings depend on the domain configuration, a repopulation among domains is accompanied by an anisotropic, reflection-selective spontaneous strain at conserved magnetic intensity. The step seen from $(\bar{1},0,\bar{1})$ at $\sim 61$~K could be understood as the elastic fingerprint of a piezomagnetically coupled domain repopulation rather than of a bulk structural transition, and it may also provide a scenario for the history-dependent FCC intensity changes observed between 60.5 and 88.5~K.

Furthermore, due to the field-induced alignment of magnetic domain states, an enhanced primary extinction of some strong reflections such as $(0,0,1)$ and $(\bar{1},0,\bar{1})$ could reduce its integrated intensity without reducing the ordered moment. The measured sample is well aligned relative to the applied in-plane field direction, however, even with a very small misalignment at 1--2 degrees between the applied magnetic field and the sample basal-plane, a small out-of-plane $c$-axis component of the applied magnetic field at 2~T could effectively be present in the experiment. Therefore, there exists a possibility that some time-reversal domains may start to be detwinned in this temperature regime due to this $c$-axis component of the field, so this abruptly intensity drop is seen..

MnTe$_2$ orders antiferromagnetically at 86.5~K~\cite{burlet_noncollinear_1997}, but it is unlikely to be relevant to the observed FCC intensity decrease: its refined fraction is about 0.5 wt\%, and its $(0,0,1)$ magnetic peak at the used D23 wavelength is expected near $2\theta=10.67^\circ$ rather than the $\alpha$-MnTe position at $11.05^\circ$. The observed anomaly in the temperature dependence of the selected reflections is therefore intrinsic to the measured $\alpha$-MnTe sample, while its microscopic origin, including possible piezomagnetic or magnetoelastic effects, remains open, further studies along this line are clearly needed.

\section{Discussion and summary}

We have demonstrated that an in-plane magnetic field efficiently detwins the orientational domains of bulk $\alpha$-MnTe, enabling the determination of the Mn moment direction $[1,1,0]$ in a single crystal using unpolarized neutron diffraction.
Table~\ref{tab:comparison} places our results in the context of the existing domain-control studies on $\alpha$-MnTe.

The field-driven detwinning approach demonstrated here offers several practical advantages.
First, it is a non-contact method, fully compatible with standard cryomagnet sample environments and enabling simultaneous application of multiple experimental probes (e.g., neutron scattering, transport, optics).
Second, the required field of $\sim 2$~T is readily accessible in most laboratory and large-scale-facility environments, in contrast to strain cells which require precise mechanical alignment and typically occupy significant space in the sample zone.
Third, the near-complete detwinning efficiency achieved at a modest field indicates that the in-plane anisotropy barriers in bulk MnTe are very small: estimates from the spin-flop field place the in-plane anisotropy constant at only $K \approx 0.2$--$1.4~\mu$eV~\cite{kriegner_magnetic_2017, mazin_altermagnetism_2023}, so that field-based domain control is both efficient and energetically economical.

We caution, however, that unpolarized neutron diffraction is insensitive to the relative population of the two antiphase ($180^\circ$) domains of the selected orientation.
The in-plane-field-prepared state should be regarded as an \textit{orientational-domain-free} state rather than a fully single-domain state.
Resolving the residual antiphase domain structure will require polarized neutron probes, and an oblique-field geometry, which addresses both domain types, appears promising for preparing a genuinely single-domain state~\cite{liu_altermagnetic_order_2026}.

The $[1,1,0]$ moment direction determined under our field-driven condition raises an intriguing question: is this direction the zero-field easy axis, or is it selected by the $[\bar{1},1,0]$ applied field through the Zeeman coupling?
Several observations indicate that the in-plane easy axis of MnTe is temperature dependent and only weakly pinned.
$\mu$SR measurements suggest that the easy axis lies along the $[1,1,0]$ direction above $\sim 250$~K and rotates by $\sim 30^\circ$, toward the $[1, 0, 0]$ direction, below that temperature~\cite{hicken_anomalous_2025}, while magnetization measurements reveal a further $90^\circ$ rotation of the easy magnetization axis concomitant with the appearance of spontaneous magnetization below $\sim 81$~K~\cite{orlova_magnetization_2025}. High-field electron spin resonance along the easy axis resolves a single, narrow pseudo-Goldstone resonance, indicating a nearly isotropic in-plane energy landscape~\cite{povarov_pseudo-goldstone_2026}.
Correspondingly, most previous studies assume a zero-field moment orientation along the $[1, 0, 0]$ direction~\cite{huynh_intrinsic_2026,kriegner_magnetic_2017,mazin_altermagnetism_2023,liu_strain-tunable_2026,zhang_li-doped_2025}, and an in-plane field has been shown to reorient the N\'eel vector continuously~\cite{weber_magnetic_2026}.
Within this picture, our 2~K, 2~T measurement selects the $[1,1,0]$ orientation through the Zeeman coupling, and the weak in-plane anisotropy makes a definitive identification of the zero-field easy axis challenging without zero-field measurements on a pre-detwinned crystal.

Whether the field-induced detwinning proceeds by discrete repopulation of the three orientational domain states or by continuous reorientation of the N\'eel vector within each domain likewise remains open, since both scenarios transfer magnetic intensity among the symmetry-related reflections and are indistinguishable in our integrated-intensity data. Complementary magnetostriction, NMR and inelastic-neutron measurements indicate that in-plane fields depopulate unfavorable orientations and continuously rotate the surviving N\'eel vectors toward the field-perpendicular configuration~\cite{duan_magnetostriction_2026,weber_magnetic_2026,huynh_intrinsic_2026}, and direct imaging of the intermediate field states, for example by scanning Kerr microscopy or nitrogen-vacancy magnetometry~\cite{watanabe_magneto-optical_2026,zhou_surface_2026}, could settle this question in bulk crystals.

Regardless of these open questions, our results provide a well-defined reference state for future studies of detwinning effects and altermagnetic phenomena.
In particular, the preparation of an orientational-domain-free bulk crystal opens the door for detecting the altermagnetic signal.
Furthermore, the ability to select a specific in-plane moment direction via field orientation offers a new degree of freedom for engineering anisotropic spin transport and magneto-optical responses in this prototypical altermagnet.

\section*{Data Availability}
The data that support the findings of this study are available from the corresponding authors upon reasonable request.
The D23, D9 and OrientExpress data are archived by the Institut Laue--Langevin under Ref.~\cite{MnTe_ILL_D9}; public access is subject to the ILL data policy and current embargo.

\section*{Acknowledgments}
This project has received funding from the European Union's Horizon 2020 research and innovation programme under the Marie Sk\l{}odowska-Curie grant agreement No. 101034266.
The authors thank Nikolaos Biniskos and Dalila Bounoua for helpful discussions.
The single-crystal growth and XRD measurements were performed at JCNS-MLZ, Garching.
The single-crystal neutron diffraction experiments were performed at the D23, D9 and OrientExpress neutron instruments at the Institut Laue--Langevin (ILL), Grenoble, France. 
We thank Markos Skoulatos and Bastian Veltel for their assistance in the DC magnetisation measurements in the Physics Lab. at the Heinz Maier-Leibnitz Zentrum (MLZ), Garching.
We thank Shibabrata Nandi for the assistance in the DC magnetisation measurements at the JCNS-2, J\"ulich.




\bibliography{MnTe_Ref.bib}

\end{document}